\documentclass[12 pt,twoside,aps,prd,amsmath,amssymb,
tightenlines,showpacs,showkeys,eqsecnum]{revtex4}
\usepackage{epsfig}
\usepackage{psfig}
\usepackage{mathptmx}
\usepackage{bm}
\usepackage{graphicx}
\renewcommand{\cal}{\mathcal}

\newcommand {\ve}{\varepsilon}
\newcommand {\pr}{\partial}
\newcommand {\cG}{\cal G}
\newcommand {\cD}{\cal D}
\newcommand {\cL}{\cal L}
\newcommand {\G}{\Gamma}
\newcommand {\bg}{\bar \gamma}
\newcommand {\bp}{\bar \psi}

\newcommand {\vf}{\varphi}

\numberwithin{equation}{section}

\begin{document}
\title{Interacting scalar and spinor fields in Bianchi type I universe 
filled with magneto-fluid}
\author{Bijan Saha}
\affiliation{Laboratory of Information Technologies\\ 
Joint Institute for Nuclear Research, Dubna\\ 
141980 Dubna, Moscow region, Russia} 
\email{saha@thsun1.jinr.ru}

\begin{abstract}
Self-consistent system of spinor, scalar and BI gravitational 
fields in presence of magneto-fluid and $\Lambda$-term is considered. 
Assuming that the expansion of the BI universe is proportional to
the $\sigma_1^1$ component of the shear tensor, exact solutions
for the metric functions, as well as for scalar and spinor fields are 
obtained. For a non-positive $\Lambda$ the initially anisotropic
space-time becomes isotropic one in the process of expansion, whereas,
for $\Lambda > 0$ an oscillatory mode of expansion of the BI model
occurs.
\end{abstract}

              

\maketitle

\bigskip

			\section{Introduction}

The discovery of the cosmic microwave radiation has stimulated a growing
interest in anisotropic, general-relativistic cosmological models of the
universe. The choice of anisotropic cosmological models in the system of
Einstein field equation enable us to study the early day universe, which
had an anisotropic phase that approaches an isotropic one~\cite{misner}.
Bianchi type I (BI) cosmological models that are anisotropic 
homogeneous universes play an important role in understanding essential
features of the universe, such as formation of galaxies during its early 
stage of evolution. An LRS BI model containing a magnetic field directed
along one axis with a barotropic fluid was investigated by Thorne
\cite{thorne}. Jacobs \cite{jacobs68,jacobs69} investigated BI models
with magnetic field satisfying a barotropic equation of state. Bali
\cite{bali} studied the behavior of the magnetic field in a BI universe
for perfect fluid distribution. 

In this paper we study the self-consistent system of spinor, scalar and
BI gravitational fields in presence of magneto-fluid and cosmological
constant. Solutions of Einstein equations coupled to a spinor and a 
scalar fields in BI spaces have been extensively studied by Saha and Shikin 
~\cite{sahagrg,sahajmp,sahaprd,sahal}. In the aforementioned papers, we
considered spinor field in BI universe where nonlinearity occurred either
due self-coupling or induced by an interacting massless scalar field.
Here, considering the BI universe filled with magneto-fluid we make an
attempt to study a system, where all the four fields, i.e., scalar, spinor, 
electro-magnetic and gravitational ones, play active part in the evolution
process.

	\section{Fundamental Equations and general solutions}
We choose the action of the self-consistent system of spinor, scalar and 
gravitational fields in the form
\begin{equation}
{\cal S}(g; \psi, \bp, \vf) = \int\, (R + {\cL}) \sqrt{-g} d\Omega,
\label{action}
\end{equation}
where $R$ is the Ricci scalar and $L$ is the spinor and scalar field 
Lagrangian density chosen in the form\cite{sahagrg} 
\begin{equation} 
{\cL}= \frac{i}{2} 
\biggl[\bp \gamma^{\mu} \nabla_{\mu} \psi- \nabla_{\mu} \bar 
\psi \gamma^{\mu} \psi \biggr] - m\bp \psi +
\frac{1}{2} \vf_{,\alpha}\vf^{,\alpha} (1 + \lambda F).
\label{lag} 
\end{equation} 
Here $\lambda$ is the coupling constant and $F$ is some 
arbitrary functions of invariants generated from the real bilinear 
forms of a spinor field. We choose $F$ to be the 
function of $I =S^2 = (\bp \psi)^2$ and 
$J = P^2 = (i \bp \gamma^5 \psi)^2$, i.e., $F = F(I, J)$, 
that describes the nonlinearity in the most general of its 
form~\cite{sahaprd}. As one sees, for $\lambda = 0$ we have the
system with minimal coupling. 

The gravitational field in our
case is given by a Bianchi type I (BI) metric in the form 
\begin{equation} 
ds^2 = a_0^2 (dx^0)^2 - a_1^2 (dx^1)^2 - a_2^2 (dx^2)^2 - a_3^2 (dx^3)^2, 
\label{BI}
\end{equation}
with $a_0 = 1$, $x^0 = c t$ and $c = 1$. The metric functions
$a_i$ $(i=1,2,3)$ are the functions of time $t$ only.

Variation of \eqref{action} with respect to spinor field $\psi\,(\bp)$
gives nonlinear spinor field equations
\begin{subequations}
\label{speq}
\begin{eqnarray}
i\gamma^\mu \nabla_\mu \psi - m \psi + {\cD} \psi + 
{\cG} i \gamma^5 \psi &=&0, \label{speq1} \\
i \nabla_\mu \bp \gamma^\mu +  m \bp - {\cD} \bp - 
{\cG} i \bp \gamma^5 &=& 0, \label{speq2}
\end{eqnarray}
\end{subequations}
where we denote
$$ {\cD} =  \lambda S \vf_{,\alpha}\vf^{,\alpha} {\pr F}/{\pr I}, 
\quad
{\cG} =  \lambda P \vf_{,\alpha}\vf^{,\alpha} {\pr F}/{\pr J},$$
whereas, variation of \eqref{action} with respect to scalar field
yields the following scalar field equation
\begin{equation}
\frac{1}{\sqrt{-g}} \frac{\pr}{\pr x^\nu} \Bigl(\sqrt{-g} g^{\nu\mu}
(1 + \lambda F) \vf_{,\mu}\Bigr) 
= 0. \label{scfe}
\end{equation}

Varying \eqref{action} with respect to metric tensor $g_{\mu\nu}$ 
one finds the gravitational field equation which in account of cosmological 
constant $\Lambda$ has the form 
\begin{subequations}
\label{BID}
\begin{eqnarray}
\frac{\ddot a_2}{a_2} +\frac{\ddot a_3}{a_3} + \frac{\dot a_2}{a_2}\frac{\dot 
a_3}{a_3}&=&  \kappa T_{1}^{1} -\Lambda,\label{11}\\
\frac{\ddot a_3}{a_3} +\frac{\ddot a_1}{a_1} + \frac{\dot a_3}{a_3}\frac{\dot 
a_1}{a_1}&=&  \kappa T_{2}^{2} - \Lambda,\label{22}\\
\frac{\ddot a_1}{a_1} +\frac{\ddot a_2}{a_2} + \frac{\dot a_1}{a_1}\frac{\dot 
a_2}{a_2}&=&  \kappa T_{3}^{3} - \Lambda,\label{33}\\
\frac{\dot a_1}{a_1}\frac{\dot a_2}{a_2} +\frac{\dot a_2}{a_2}\frac{\dot 
a_3}{a_3}+\frac{\dot a_3}{a_3}\frac{\dot a_1}{a_1}&=&  \kappa T_{0}^{0} - 
\Lambda.
\label{00}
\end{eqnarray}
\end{subequations}
Here $\kappa$ is the Einstein gravitational constant and over-dot means 
differentiation with respect to $t$. The energy-momentum tensor of the 
system is given by
\begin{eqnarray}
T_{\mu}^{\rho} &=& \frac{i}{4} g^{\rho\nu} \biggl(\bp \gamma_\mu 
\nabla_\nu \psi + \bp \gamma_\nu \nabla_\mu \psi - \nabla_\mu \bar 
\psi \gamma_\nu \psi - \nabla_\nu \bp \gamma_\mu \psi \biggr)  \label{tem}\\
& & + (1 - \lambda F)  \vf_{,\mu}\vf^{,\rho} - \delta_{\mu}^{\rho} {\cL}
+ T_{{\rm m}\,\mu}^{\,\,\,\nu}. \nonumber
\end{eqnarray}
The energy-momentum tensor of the magneto-fluid is chosen to be
\begin{equation} 
T_{\mu\,(m)}^{\nu} = (\ve + p) v_\mu v^\nu - p \delta_\mu^\nu + E_\mu^\nu,
\label{imperfl}
\end{equation}
where $E_{\mu\nu}$ is the electro-magnetic field given by 
Lichnerowich \cite{lich}
\begin{equation}
E_\mu^\nu = {\bar \mu} \Bigl[ |h|^2 \Bigl(u_\mu u^\nu - \frac{1}{2}
\delta_\mu^\nu\Bigr) - h_\mu h^\nu \Bigr].
\label{lichn}
\end{equation}
Here $u^\mu$ is the flow vector satisfying
\begin{equation}
g_{\mu\nu} u^\mu u^\nu = 1,
\label{scprod}
\end{equation}
$\bar \mu$ is the magnetic permeability and $h_\mu$ is the magnetic 
flux vector defined by
\begin{equation}
h_\mu = \frac{1}{\bar \mu} \star F_{\nu \mu} u^\nu,
\label{magflux}
\end{equation}
where $\star F_{\mu\nu}$ is the dual electro-magnetic field tensor defined 
as 
\begin{equation}
\star F_{\mu \nu} = \frac{\sqrt{-g}}{2} \epsilon_{\mu \nu \alpha \beta}
F^{\alpha \beta}.
\label{dualt}
\end{equation}
Here $F^{\alpha \beta}$ is the electro-magnetic field tensor and
$\epsilon_{\mu \nu \alpha \beta}$ is the totally anti-symmetric Levi-Civita
tensor with $\epsilon_{0123} = +1$. Here the comoving coordinates are
taken to be $u^0 = 1,\, u^1 = u^2 = u^3 = 0$. We choose the incident
magnetic field to be in the direction of $x$-axis so that the magnetic
flux vector has only one nontrivial component, namely $h_1 \ne 0.$ In view
of the aforementioned assumption from \eqref{magflux} we obtain 
$F_{12} = F_{13} = 0.$
We also assume that the conductivity of the fluid is infinite. This leads to
$F_{01} = F_{02} = F_{03} = 0$. Thus we have only one non-vanishing component
of $F_{\mu \nu}$ which is $F_{23}.$ 
Then from the first set of Maxwell equation
\begin{equation}
F_{\mu\nu;\beta} + F_{\nu\beta;\mu} + F_{\beta \mu; \nu} = 0,
\label{maxe}
\end{equation}
where the semicolon stands for covariant derivative, one finds
\begin{equation}
F_{23} = {\cal I}, \quad {\cal I} = {\rm const.}
\label{f23}
\end{equation} 
Then from \eqref{magflux} in account of \eqref{dualt}  one finds
\begin{equation}
h_1 = \frac{a_1 {\cal I}}{{\bar \mu} a_2 a_3}.
\label{h1}
\end{equation}
Finally, for $E_\mu^\nu$ we find the following non-trivial components
\begin{equation}
E_0^0 = E_1^1 = - E_2^2 = - E_3^3 = \frac{{\cal I}^2}
{2 {\bar \mu} a_2^2 a_3^2}.
\label{E}
\end{equation}

In \eqref{imperfl} $\ve$ and $p$ are the energy and pressure of the
fluid, respectively. In this note we assume that the equation of state
\begin{equation}
p = \zeta \ve
\label{stateq}
\end{equation}
holds. Here $\zeta$ varies between the
interval $0\,\le\, \zeta\,\le\,1$, whereas $\zeta\,=\,0$ describes
the dust Universe, $\zeta\,=\,\frac{1}{3}$ presents radiation Universe,
$\frac{1}{3}\,<\,\zeta\,<\,1$ ascribes hard Universe and $\zeta\,=\,1$
corresponds to the stiff matter. The Dirac matrices $\gamma_\mu(x)$
of curve space-time are connected with those of Mincowski space as
\begin{equation}
\gamma^\mu = \bg^\mu/a_\mu, \quad \gamma_\mu = \bg a_\mu, \quad
\mu = 0,1,2,3. 
\label{gamma}
\end{equation}

In the Eqs. \eqref{speq} and \eqref{tem} $\nabla_\mu$ is the covariant
derivatives acting on a spinor field as ~\cite{brill}
\begin{equation}
\nabla_\mu \psi = \frac{\partial \psi}{\partial x^\mu} -\G_\mu \psi, \quad
\nabla_\mu \bp = \frac{\partial \bp}{\partial x^\mu} + \bp \G_\mu, 
\label{cvd}
\end{equation}
where $\G_\mu$ are the Fock-Ivanenko spinor connection coefficients
defined by
\begin{equation}
\G_\mu = \frac{1}{4} \gamma^\sigma \Bigl(\G_{\mu \sigma}^{\nu} \gamma_{\nu}
- \partial_{\mu} \gamma_{\sigma}\Bigr).  
\label{fock}
\end{equation}  
For the metric \eqref{BI} one has the following components
of the spinor connection coefficients 
\begin{equation}  
\G_\mu = (1/2){\dot a_\mu} \bg^\mu \bg^0.
\label{safc}
\end{equation}
 
We study the space-independent solutions to the spinor 
and scalar field Eqs. \eqref{speq} and \eqref{scfe} so that 
$\psi=\psi(t)$ and $\vf = \vf(t)$.
defining
\begin{equation}
\tau = a_0 a_1 a_2 a_3 = \sqrt{-g}
\label{taudef}
\end{equation}
from \eqref{scfe} for the scalar field  we have
\begin{equation}
\vf = C \int [\tau (1 + \lambda F)]^{-1} dt.
\label{sfsol}
\end{equation}

Setting $V_j(t) = \sqrt{\tau} \psi_j(t), \quad j=1,2,3,4,$ in view
of \eqref{cvd} and \eqref{safc} 
from \eqref{speq1} one deduces the following system of equations:  
\begin{subequations}
\label{V}
\begin{eqnarray} 
\dot{V}_{1} + i (m - {\cD}) V_{1} - {\cG} V_{3} &=& 0, \\
\dot{V}_{2} + i (m - {\cD}) V_{2} - {\cG} V_{4} &=& 0, \\
\dot{V}_{3} - i (m - {\cD}) V_{3} + {\cG} V_{1} &=& 0, \\
\dot{V}_{4} - i (m - {\cD}) V_{4} + {\cG} V_{2} &=& 0. 
\end{eqnarray} 
\end{subequations}

From \eqref{speq1} we also write the equations for the bilinear spinor 
forms $S,\quad P$ and $A^0 = \bp \bg^5 \bg^0 \psi$
\begin{subequations}
\label{inv}
\begin{eqnarray}
{\dot S_0} - 2 {\cG}\, A_0^0 &=& 0, \label{S0}\\
{\dot P_0} - 2 (m - {\cD})\, A_0^0 &=& 0, \label{P0}\\
{\dot A_0^0} + 2 (m - {\cD})\, P_0 + 2 {\cG} S_0 &=& 0, \label{A0} 
\end{eqnarray}
\end{subequations}
where $Q_0 = \tau Q$, leading to the relation
$S^2 + P^2 + (A^0)^2 =  C^2/ \tau^2, \qquad C^2 = {\rm const.}$
As one sees, for $F=F(I)$ \eqref{S0} gives $S = C_0/\tau$,
whereas for the massless spinor field with $F=F(J)$ \eqref{P0}
yields $P=D_0/\tau$. In view of it for $F=F(I)$ we obtain the
following expression for the components of spinor field
\begin{eqnarray} 
\psi_1(t) &=& C_1 \tau^{-1/2} e^{-i\beta}, \quad
\psi_2(t) = C_2 \tau^{-1/2} e^{-i\beta},  \nonumber\\
\label{spef}\\
\psi_3(t) &=& C_3 \tau^{-1/2} e^{i\beta}, \quad
\psi_4(t) = C_4 \tau^{-1/2} e^{i\beta},
\nonumber
\end{eqnarray} 
with $C_i$ being the integration constants and
are related to $C_0$ as 
$C_0 = C_{1}^{2} + C_{2}^{2} - C_{3}^{2} - C_{4}^{2}.$ Here
$\beta = \int(m - {\cD}) dt$. In case of $F=F(J)$ for the massless 
spinor field we get 
\begin{eqnarray}
\psi_1 &=& \tau^{-1/2} \bigl(D_1 e^{i \sigma} + 
iD_3 e^{-i\sigma}\bigr), \quad
\psi_2 = \tau^{-1/2} \bigl(D_2 e^{i \sigma} + 
iD_4 e^{-i\sigma}\bigr), \nonumber \\
\label{J}\\
\psi_3 &=& \tau^{-1/2} \bigl(iD_1 e^{i \sigma} + 
D_3 e^{-i \sigma}\bigr),\quad
\psi_4 = \tau^{-1/2} \bigl(iD_2 e^{i \sigma} + 
D_4 e^{-i\sigma}\bigr). \nonumber
\end{eqnarray} 
The integration constants $D_i$
are connected to $D_0$ by
$D_0=2\,(D_{1}^{2} + D_{2}^{2} - D_{3}^{2} -D_{4}^{2}).$
Here we set $\sigma = \int {\cG} dt$. 

Once the spinor functions are known explicitly, one can write the components 
of spinor current
$j^\mu = \bp \gamma^\mu \psi$,
the charge density of spinor field 
$\varrho = (j_0\cdot j^0)^{1/2}$,
the total charge of spinor field 
$Q = \int\limits_{-\infty}^{\infty} \varrho \sqrt{-^3 g} dx dy dz,$
the components of spin tensor
$
S^{\mu\nu,\epsilon} = \frac{1}{4}\bp \bigl\{\gamma^\epsilon
\sigma^{\mu\nu}+\sigma^{\mu\nu}\gamma^\epsilon\bigr\} \psi$
and other physical quantities.

Let us now solve the Einstein equations. In doing so we first write the 
expressions for the components of the energy-momentum tensor explicitly:
\begin{subequations}
\label{total}
\begin{eqnarray}
T_{0}^{0} &=& mS + C^2/2\tau^2 (1+\lambda F) + \ve + 
\frac{{\cal I}^2}{2 {\bar \mu} a_2^2 a_3^2}, \label{t00}\\
T_{1}^{1} &=& {\cD} S + {\cG} P - C^2/2\tau^2 (1+\lambda F) - p
+\frac{{\cal I}^2}{2 {\bar \mu} a_2^2 a_3^2}, \label{t11}\\
T_{2}^{2} &=& {\cD} S + {\cG} P - C^2/2\tau^2 (1+\lambda F) - p
-\frac{{\cal I}^2}{2 {\bar \mu} a_2^2 a_3^2}, \label{t22}\\
T_{3}^{3} &=& {\cD} S + {\cG} P - C^2/2\tau^2 (1+\lambda F) - p 
-\frac{{\cal I}^2}{2 {\bar \mu} a_2^2 a_3^2}, \label{t33}\\
\end{eqnarray}
\end{subequations}
In view of $T_{2}^{2} = T_{3}^{3}$ from \eqref{22}, \eqref{33}
we find 
\begin{equation} 
a_2 = a_3 D {\rm exp}\Bigl(X \int \frac{dt}{\tau}\Bigr), 
\label{b/c}
\end{equation}
with the constants of integration $D$ and $X$ being
integration constants.

Following Bali \cite{bali} let us assume that the expansion ($\theta$)
in the model is proportional to the eigenvalue $\sigma_1^1$ of the shear
tensor $\sigma_\mu^\nu$. Since for the BI space-time
\begin{eqnarray}
\theta &=& \frac{\dot a_1}{a_1}+\frac{\dot a_2}{a_2}+\frac{\dot a_3}{a_3}, \\
\sigma_1^1 &=& - \frac{1}{3}\Bigl(4\frac{\dot a_1}{a_1}+\frac{\dot a_2}{a_2}+
\frac{\dot a_3}{a_3}\Bigr), 
\end{eqnarray}
the aforementioned condition leads to 
\begin{equation}
a_1 = \bigl(a_2 a_3 \bigr)^N,
\label{a=bc}
\end{equation}
with $N$ being the proportionality constant. 

In account of \eqref{taudef} from \eqref{b/c} and \eqref{a=bc} after some
manipulation for the metric functions one finds 
\begin{subequations}
\label{abc}
\begin{eqnarray}
a_1 &=& \tau^{N/(N+1)}, \label{a1}\\
a_2 &=& \sqrt{D}\, \tau^{1/2(N+1)}\, {\rm exp} \Bigl[\frac{X}{2} \int 
\frac{dt}{\tau}\Bigr], \label{a2}\\ 
a_3 &=& \frac{1}{\sqrt{D}}\, \tau^{1/2(N+1)}\, {\rm exp} \Bigl[-\frac{X}{2} 
\int \frac{dt}{\tau}\Bigr]. \label{a3}
\end{eqnarray}
\end{subequations} 
As one sees from \eqref{abc} for $\tau \sim t^n$
with $n > 1$ the exponent tends to unity at large $t$. In this case the 
anisotropic model becomes isotropic one iff $D = 1$ and $N = 1/2$. 
Let us also write the invariants of gravitational field. They are
the Ricci scalar $I_1 = R \approx 1/\tau^2$,
$I_2 = R_{\mu\nu}R^{\mu\nu} \equiv R_{\mu}^{\nu} R_{\nu}^{\mu}
\approx 1/\tau^4$ and the Kretschmann scalar
$I_3 = R_{\alpha\beta\mu\nu}R^{\alpha\beta\mu\nu} \approx 1/\tau^4$.
As we see, the space-time becomes singular at a point where $\tau = 0$,
as well as the scalar and spinor fields. Thus we see, all the functions
in question are expressed via $\tau$. In what follows, we write the
equation for $\tau$ and study it in details.  

Summation of Einstein Eqs. \eqref{11}, \eqref{22}, \eqref{33} and 
\eqref{00} multiplied by 3 gives
\begin{equation}
\frac{\ddot 
\tau}{\tau}= \frac{3}{2}\kappa \Bigl(mS + {\cD} S + {\cG} P + \ve -p
+ \frac{2 {\cal I}^2}{3 {\bar \mu} (a_2 a_3)^2} \Bigr) - 3 \Lambda. 
\label{dtau1}
\end{equation} 
For the right hand side of Eq. \eqref{dtau1} to be a function of $\tau$
only, the solution to this equation is well known \cite{kamke}.
In what follows we study this equation for some concrete form of $F$.
In doing so 
let us demand the energy-momentum to be conserved, i.e.,
$T_{\mu;\nu}^{\nu} = 0$, which in our case takes the form
\begin{equation}
\frac{1}{\tau}\bigl(\tau T_0^0\bigr)^{\cdot} - \frac{\dot a_1}{a_1} T_1^1
-\frac{\dot a_2}{a_2} T_2^2  - \frac{\dot a_3}{a_3} T_3^3 = 0.
\label{emcon}
\end{equation}
In account of the equation of state \eqref{stateq} and 
$$(m -{\cD}) \dot{S}_0 - {\cG} \dot{P}_0 = 0$$
which follows from \eqref{inv}, after a little manipulation from
\eqref{emcon} we obtain  
\begin{equation}
\ve = {\ve_0}/{\tau^{1+\zeta}},\quad 
p = {\zeta \ve_0}/{\tau^{1+\zeta}}.
\label{vep}
\end{equation}

Let us recall that we consider $F$ as a function of $I$, $J$ or $I\pm J$.
If we choose $F=F(I)$, then setting $m=0$ we come to the analogical equation 
corresponding to the massless spinor case with $F=F(J)$ or $F=F(I \pm J)$,
whereas setting $\lambda = 0$ we have the system with minimal coupling. 
Under this assumption from \eqref{inv} one finds 
\begin{equation}
S = \frac{C_0}{\tau}.
\label{stau}
\end{equation}
In view of \eqref{stau}, \eqref{vep} and the fact that 
$a_2 a_3 = \tau^{1/(N+1)}$, Eq. \eqref{dtau1} can be rewritten as
\begin{equation}
{\ddot \tau}= \frac{3}{2}\kappa \Bigl(m C_0 + {\cD} C_0 +  \ve_0 (1 - \zeta)/
\tau^{\zeta} + \frac{2 {\cal I}^2}{3 {\bar \mu}}\tau^{(N-1)/(N+1)}
 \Bigr) - 3 \Lambda. 
\label{dtaui}
\end{equation} 
Recalling the definition of ${\cD}$ the solution to Eq. \eqref{dtaui}
can be written in quadrature
\begin{equation}
\int \frac{d \tau}{\sqrt{\kappa\Bigl(m C_0 \tau + C^2/2(1 + \lambda F)
+ \ve_0 \tau^{1 - \zeta} + ((N+1){\cal I}^2/3 {\bar \mu} N) 
\tau^{2N/(N+1)}\Bigr) - \Lambda \tau^2 + E }} \,= \, \sqrt{3}\, t,
\label{quadra}
\end{equation}
with $E$ being some integration constant. It should be mentioned that
being the volume-scale $\tau$ is non-negative. At the points where 
$\tau = 0$ there occurs space-time singularity. On the other hand,  
the radical in \eqref{quadra} should be positive.
This fact leads to the conclusion that for $\Lambda > 0$ the value of
$\tau$ is bound from above as well, giving rise to an oscillatory mode
of expansion of the BI universe. For a non-positive $\Lambda$, we
have picture with fast expanding $\tau$ with time.    

			\section{Conclusions}	

A self-consistent system of spinor, scalar and gravitation fields has been 
studied in presence of magneto-fluid and cosmological term $\Lambda$. 
With the presence of $F_{23}$ component of electro-magnetic field tensor, the 
system can be viewed as one where all the four fields, i.e., scalar, 
electro-magnetic, spinor and gravitational ones, are taken into 
consideration. Assuming that the expansion of the BI space-time is 
proportional to the $\sigma_1^1$ component of the shear tensor, solutions 
for the metric functions $a_i(t)$ are obtained explicitly in terms of 
volume-scale $\tau$. Expressions for the scalar and spinor fields are 
also obtained in terms of $\tau$. For the non-positive $\Lambda$ we 
obtain exponentially expanding BI universe, which means the initially
anisotropic space-time becomes isotropic one in the process of expansion.
For a positive $\Lambda$ an oscillatory mode of expansion takes place.
Choosing the integration constant $E$ and initial value value of $\tau$
it is possible to obtain solutions those are regular everywhere.
A detailed numerical study of the Eq. \eqref{dtaui} we plan to perform
in short. 
\begin{acknowledgments}

\end{acknowledgments}



\begin{thebibliography}{99}

\bibitem{misner} Misner C.W. {\it Astrophysical Journal} 
{\bf 151}, 431 (1968).

\bibitem{thorne} Thorne K.S. {\it Astrophysical Journal} 
{\bf 148}, 51 (1967).

\bibitem{jacobs68} Jacobs K.C. {\it Astrophysical Journal}
{\bf 153}, 661 (1968).

\bibitem{jacobs69} Jacobs K.C. {\it Astrophysical Journal}
{\bf 155}, 379 (1969).

\bibitem{bali} Bali R. {\it International Journal of Theoretical Physics} 
{\bf 25}, 755 (1986).

\bibitem{sahagrg} Saha B., Shikin G.N. 
{\it General Relativity and Gravitation} {\bf 29}, 1099 (1997). 

\bibitem{sahajmp} Saha B., Shikin G.N.
{\it Journal Mathematical Physics} {\bf 38}, 5305 (1997).

\bibitem{sahaprd} Bijan Saha
{\it Physical Review D} {\bf 64}, 123501 (2001).

\bibitem{sahal} Bijan Saha 
{\it Modern Physics Letters A} {\bf 16}, 1287 (2001). 

\bibitem{lich} Lichnerowicz A.   
{\it Relativistic Hydrodynamics and Magnetohydrodynamics}, (Benjamin,
New York)

\bibitem{brill} Brill D. and Wheeler J. 
{\it Review of Modern Physics} {\bf 29}, 465 (1957).  

\bibitem{kamke} E. Kamke, {\it Differentialgleichungen losungsmethoden
und losungen} (Akademische Verlagsgesellschaft, Leipzig, 1957). 

\end{thebibliography}
\end{document}